# Spatiotemporally decoupled super-oscillatory microscopy enables label-free reconstruction of human ciliary waveforms

Ning Xu[1,2], Sarah E. Bohndiek*[1,3], Graham Spicer[1], Mingjun Jiang[4,5], Qixia Wang[2], Baoping Xu[4,5], Qiaofeng Tan[2], and Calum Williams*[1,6]

*[1]Department of Physics, Cavendish Laboratory, University of Cambridge, JJ Thomson Avenue, Cambridge, CB3 0HE, UK*

*[2]State Key Laboratory of Precision Measurement Technology and Instruments, Department of Precision Instrument, Tsinghua University, Beijing 100084, China*

*[3]Cancer Research UK Cambridge Institute, University of Cambridge, Robinson Way, Cambridge, CB2 0RE, UK*

*[4]Department of Respiratory Disease, Beijing Children's Hospital, Capital Medical University, National Center for Children's Health, Beijing 100045, China.*

*[5]China National Clinical Research Center of Respiratory Diseases, Beijing 100045, China*

*[6]Department of Physics and Astronomy, University of Exeter, Stocker Road, Exeter, EX4 4QL, United Kingdom*

*Corresponding author: seb53@cam.ac.uk, C.Williams15@exeter.ac.uk

## Abstract

Imaging native motile cilia requires label-free contrast, sub-diffraction spatial discrimination, and millisecond temporal sampling. We introduce Super-Oscillatory Label-free Inertia-free Scanning (SOLIS) microscopy, in which a static multilevel diffractive optical element generates compressed super-oscillatory probes and a high-speed digital micromirror device addresses the probes without translated-stage motion. A complete 25-position scan produces one acquisition-window-averaged map every 3.58 ms, corresponding to 279 reconstructed fps. On nanofabricated line-pair targets, SOLIS retained measurable label-free contrast at a 253 nm period and resolved representative modulation at 270 nm. In differentiated primary human nasal epithelial cultures, SOLIS distinguished adjacent ciliary maxima separated by 346 nm, below the 431 nm Rayleigh reference, and acquired 14–16 reconstructed samples per observed beat cycle. Tip-centroid tracking recovered closed, asymmetric two-dimensional trajectories with a median cycle-to-cycle localisation precision of 63.5 nm across 20 cilia. In a proof-of-principle perturbation, Dynarrestin reduced culture-level ciliary beat frequency from 19.2 ± 0.6 to 6.6 ± 0.5 Hz and projected stroke amplitude from 4.8 ± 0.3 to 2.3 ± 0.2 μm. By separating spatial compression from temporal addressing, SOLIS extends super-oscillatory microscopy to label-free, millisecond-scale analysis of human ciliary kinematics.

## Introduction

Motile cilia are active cellular oscillators whose coordinated beat waveforms drive mucociliary clearance and encode essential information about epithelial function [1-3]. Resolving their native motion requires an unusual combination of optical capabilities, namely sub-diffraction spatial discrimination to distinguish closely spaced motile cilia and adjacent ciliary features, millisecond-scale temporal sampling to preserve asymmetric effective and recovery strokes, and label-free contrast to avoid perturbing native ciliary motion [4, 5]. Conventional optical imaging can capture ciliary beat frequency (CBF) at high frame rates, but it lacks the spatial resolution needed to resolve waveform-level structural motion. Dedicated high-speed video microscopy routinely samples ciliary beating at 100-1,000 frames per second [6, 7] and is label-free, but remains diffraction-limited and cannot resolve structural detail below the diffraction limit. Super-resolution microscopy provides nanoscale spatial information but remains constrained for fast, native ciliary dynamics by mechanical scanning inertia, modulation-bandwidth limits, fluorescence labelling requirements or optical-dose constraints [8-10]. Ciliary kinematic imaging requires sub-diffraction spatial sampling, dense temporal readout, and label-free contrast in the same experiment.

Super-oscillatory optics provide one route towards far-field sub-diffraction imaging. By employing phase-only diffractive optical elements (DOEs), light fields can be engineered to oscillate locally faster than their highest Fourier components, generating customised far-field super-resolution spot arrays [11-14]. This concept previously enabled adjustable super-resolution microscopy through DOE-engineered focal fields [15-17]. This first-generation design relied on motorised stage scanning. The physical mass of such actuators imposes an inertia-limited scanning speed, which makes the approach unsuitable for waveform-level reconstruction of rapidly beating cilia. When temporal sampling is insufficient to resolve the relevant frequency content and waveform phases of the biological motion [18], undersampling produces motion blur and spatiotemporal aliasing, obscuring the true trajectory of the moving structure [19].

To remove mechanical scanning inertia, we recently developed a mechanical-scan-free architecture for multicolour fluorescence super-resolution imaging [20]. This approach demonstrated that optical scanning can be decoupled from physical sample translation. However, its temporal performance remained limited by the liquid-crystal relaxation time of spatial light modulators (SLMs) [21]. Maintaining high-fidelity 2π phase modulation for super-oscillatory wavefronts restricted the effective updating rate to the order of ~20 Hz (~50 ms per pattern), making the system more suitable for fixed-cell or slowly varying samples than for fast periodic motion [22, 23]. The liquid-crystal response time therefore limits the achievable update rate. High-speed digital micromirror devices (DMDs) offer a complementary capability, providing programmable binary holograms that steer light at kHz refresh rates [24-27]. Binary-amplitude DMDs cannot by themselves generate the high-fidelity multi-level wavefront that super-oscillatory compression demands. The DMD supplies rapid addressing, while the static super-oscillatory DOE supplies probe compression. In addition, fluorescence-based implementations require exogenous labels and excitation intensities that can introduce photobleaching or phototoxicity during sustained high-speed observation [28, 29]. These limitations are especially relevant for motile cilia, whose native beat patterns can potentially be altered by optical or biochemical perturbations [30].

We introduce Super-Oscillatory Label-free Inertia-free Scanning (SOLIS) microscopy, which assigns spatial compression and temporal addressing to separate optical modules. A custom multilevel DOE statically generates super-oscillatory probe arrays, whereas a high-speed DMD digitally addresses the probes without translated-stage motion. The integrated system retains measurable contrast at a 253 nm line-pair period in a static label-free target; complementary fluorescence tests distinguish a 142 nm mitochondrial separation in fixed cells and follow mitochondrial remodelling in live cells (Supplementary Figs. 4 and 5). The principal application is label-free imaging of differentiated primary human nasal epithelial cultures at 279 reconstructed fps. Each map integrates one 3.58 ms scan and represents an acquisition-window average of ciliary motion. This temporal density recovers asymmetric effective and recovery strokes and detects pharmacologically induced changes in beat frequency and motion amplitude.

## Results

### Theoretical framework of spatiotemporally decoupled SOLIS imaging

Capturing ciliary motion requires an imaging scheme that preserves spatial discrimination while sampling the beat densely enough to retain its phase. In mechanically scanned super-resolution systems, these requirements are coupled because denser spatial sampling requires sequential acquisition, while scanner inertia and settling limit the reconstruction rate. This constraint is particularly important for human motile cilia, which typically beat at 15–20 Hz and exhibit asymmetric effective and recovery strokes.

SOLIS addresses this through a spatiotemporally decoupled optical design in which spatial field compression and temporal beam addressing are assigned to separate hardware modules. In the spatial domain, a multi-level phase-only DOE generates far-field super-oscillatory illumination probes. By modulating the pupil function $P(\mathbf{k})$, the DOE shapes the focal intensity distribution $I_{\text{focal}}(\mathbf{r})$ according to [31],

$$I_{\text{focal}}(\mathbf{r}) = \left| \int P(\mathbf{k}) \exp\left[ \mathrm{i}\Phi_{\text{DOE}}(\mathbf{k}) \right] \exp(\mathrm{i}\mathbf{k}\cdot\mathbf{r})\, d\mathbf{k} \right|^2 , \qquad (1)$$

where $\mathbf{k}$ is the wavevector and $\Phi_{\text{DOE}}(\mathbf{k})$ represents the optimised phase profile. This static modulation defines the sub-diffraction sampling kernel independently of temporal addressing. For label-free reconstruction, we use a phenomenological weak-contrast model in which the measured local intensity is parameterised by an effective two-dimensional scattering contrast. The model does not recover quantitative phase or refractive index and does not describe multiple scattering; its conditional operating range is evaluated in Supplementary Note 1.

In the temporal domain, a high-speed DMD functions as a dynamic binary hologram to scan the super-oscillatory probes across the specimen plane without mechanical sample translation. The spatiotemporal sampling function $S(\mathbf{r},t)$ of the complete dual-diffractive system can be expressed as a discrete scanning sequence [32],

$$S(\mathbf{r},t) = \sum_{n} I_{\text{focal}}(\mathbf{r}-\mathbf{d}_n)\cdot \mathrm{rect}\left[ \frac{t-n\Delta t}{\tau} \right] , \qquad (2)$$

where $\mathbf{d}_n$ is the programmable displacement vector encoded by the $n$-th DMD hologram, $\Delta t$ is the pattern-to-pattern interval, and $\tau$ is the per-pattern dwell time. Digital addressing removes the acceleration and settling associated with translated mechanical scanners. Finite physical micromirror switching remains included in the measured timing. The SOLIS architecture is summarised in Fig. 1a,b and Supplementary Fig. 2.

For the label-free ciliary experiments, one 5×5 scan comprised 25 sequential DMD patterns and was completed in 3.58 ms. The 3.58 ms cycle corresponds to 279 reconstructed fps, distinct from the raw DMD switching frequency. A 15–20 Hz beat spans 50–67 ms, so SOLIS provides approximately 14–19 acquisition windows per expected beat cycle. Each map is a 3.58 ms temporal average. SOLIS combines static structural discrimination with densely sampled tip-centroid kinematics. Each reconstructed map remains an acquisition-window average and does not represent an instantaneous full-resolution image of a moving cilium (Fig. 1c and Fig. 3).

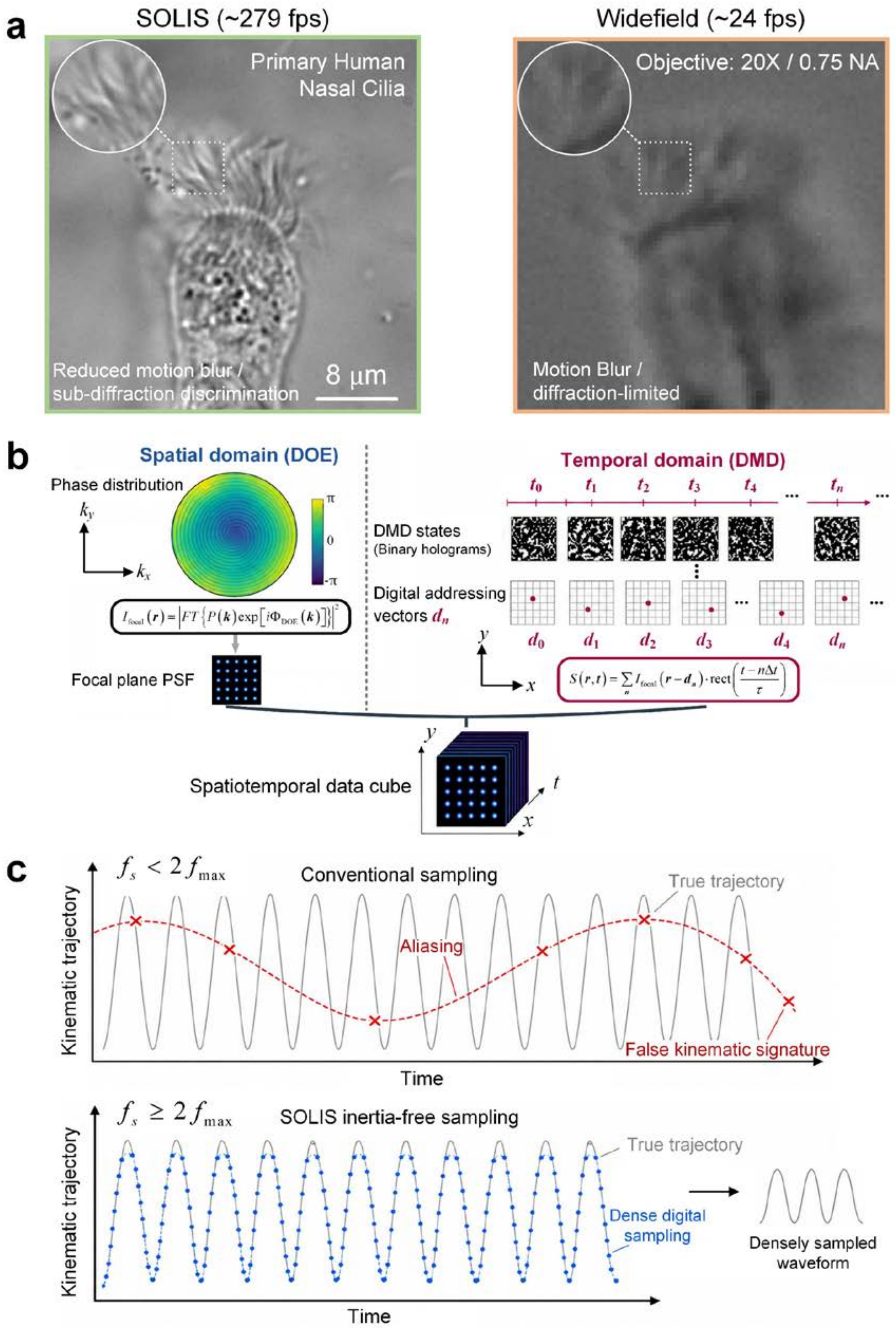

**Fig. 1. Principle and sampling requirements of SOLIS imaging. a**, Representative label-free SOLIS and widefield images of cilia in differentiated primary human nasal epithelial cultures. SOLIS images are reconstructed at approximately 279 fps; the widefield comparison is shown at approximately 24 fps. The 8 µm scale bar applies to both images. **b**, Dual-diffractive separation of spatial and temporal functions. A multilevel DOE shapes the pupil phase $\Phi_{\mathrm{DOE}}(\mathbf{k})$ and produces the focal probe array $I_{\mathrm{focal}}(\mathbf{r})$. A DMD displays binary holograms that address calibrated displacement vectors $\mathbf{d}_n$ at pattern times $t_n$. $\Delta t$ is the pattern interval and $\tau$ is the dwell time. The resulting measurements form a spatiotemporal data cube. **c**, Conceptual temporal-sampling comparison. Sparse sampling below the relevant waveform bandwidth aliases the trajectory, while dense digital sampling retains its phase evolution. The schematic illustrates sampling logic and is not a measured frequency trace.

## Spatial engineering and inertia-free optical addressing

We fabricated custom 16-level phase-only DOEs by aligned lithography and reactive-ion etching (Methods). Scanning electron microscopy confirmed the surface topology (Fig. 2a). Under 530 nm illumination, measured focal-plane distributions showed a central-lobe FWHM of 203 nm for a single super-oscillatory focus and 224 nm for the 5×5 array, compared with a 360 nm diffraction-limited FWHM (Fig. 2c). The corresponding ratios are 0.56 and 0.62, respectively. Across array configurations, intensity non-uniformity increased from 3.0% to 7.1%, and the signal-to-background ratio decreased from 8.3 to 5.3 (Supplementary Fig. 3a–d), identifying the 5×5 array as a practical compromise between parallel addressing and probe quality.

A high-speed DMD encoded binary diffraction holograms that displaced the probe array without mechanical sample translation. Calibration of the sample-plane centroids gave equal lateral increments of $\Delta x = \Delta y = 87$ nm (Fig. 2b), which oversample the 224 nm central lobe. Digital addressing avoids mechanically induced acceleration and settling; finite micromirror switching remains part of the hardware timing (Supplementary Fig. 2).

We evaluated static label-free contrast transfer with nanofabricated line-pair targets (NA = 0.75, λ = 530 nm). The conventional incoherent periodic cutoff is 353 nm (λ/2NA). Widefield contrast approached the noise floor near this cutoff, whereas SOLIS retained measurable contrast at the tested 253 nm period (Fig. 2d). Representative images show resolved modulation at 270 nm, compared with 375 nm, in the SOLIS reconstruction but not in the matched widefield image (Fig. 2e). We therefore distinguish the smallest period with measurable empirical contrast (253 nm) from the representative image-based demonstration (270 nm).

Subsequent analyses used a central-lobe FWHM of 224 nm, probe non-uniformity of 7.1%, an SBR of 5.3, a reassignment interval of 87 nm, and a full-scan time of 3.58 ms. We then evaluated the effects of sequential acquisition on dynamic tip-centroid localisation and waveform recovery.

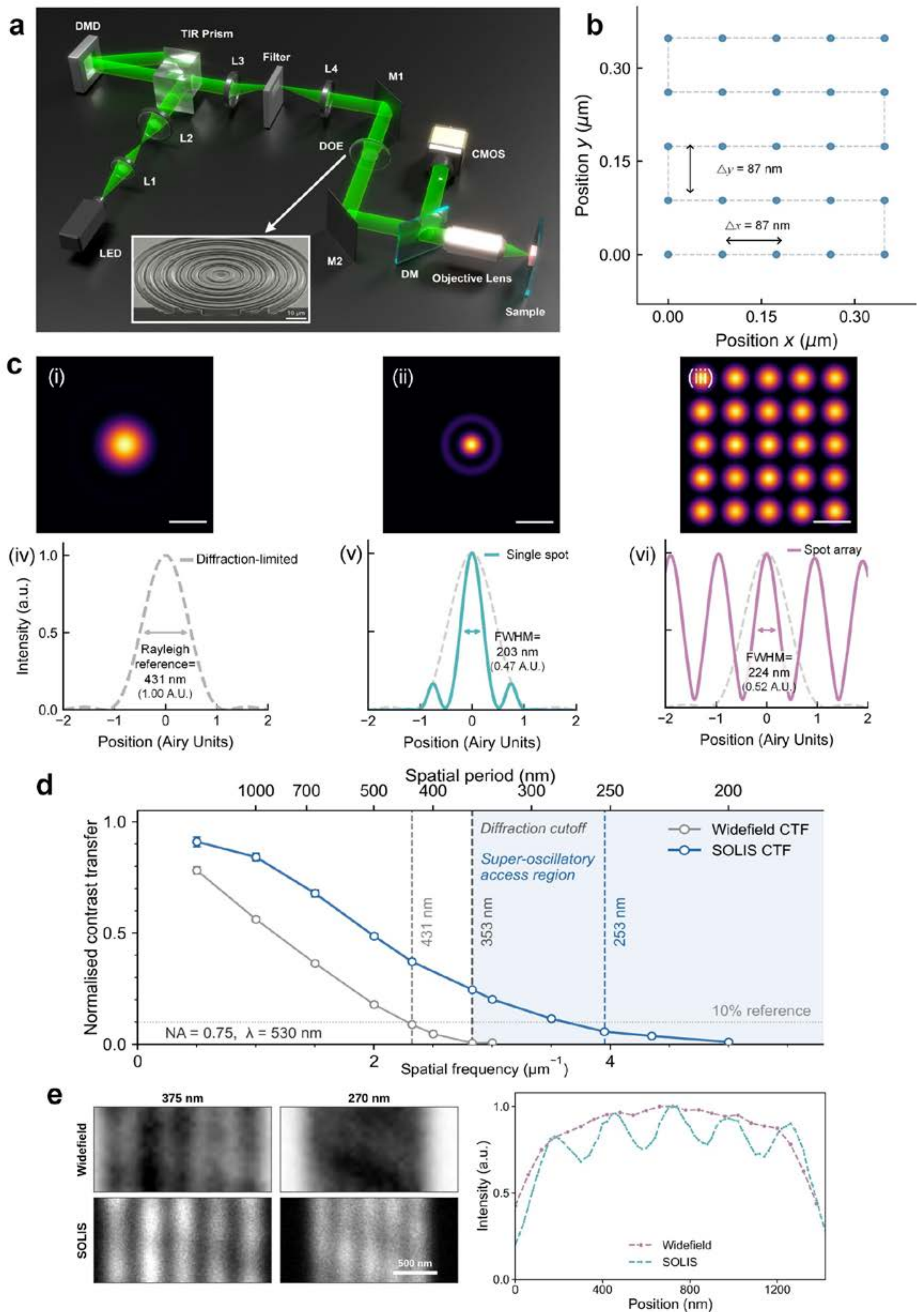

**Fig. 2. Optical implementation, digital addressing, and sub-diffraction contrast transfer of SOLIS. a**, Schematic of the SOLIS optical system. Illumination is directed onto the DMD through a total-internal-reflection prism; the selected diffraction order is relayed to the multilevel DOE, which generates the probe array at the sample plane. Scattered or emitted light is collected by the objective and recorded by a CMOS

camera. Inset, scanning electron micrograph of the fabricated 16-level DOE; scale bar, 10 μm. **b**, Sample-plane calibration of DMD addressing. The 25 programmed positions form a regular grid with $\Delta x = \Delta y$ = 87 nm. **c**, Measured focal-plane distributions of a diffraction-limited focus (i), a single super-oscillatory focus (ii), and a 5×5 array (iii), with corresponding normalised profiles (iv to vi). The displayed 431 nm reference width is the Rayleigh distance, 0.61λ/NA. The measured super-oscillatory central-lobe FWHM values are 203 and 224 nm, compared with a 360 nm diffraction-limited FWHM (NA = 0.75, λ = 530 nm). **d**, Empirical contrast transfer from nanofabricated line-pair targets. Contrast $C = \left(I_{\max} - I_{\min}\right) / \left(I_{\max} + I_{\min}\right)$ was normalised to the low spatial frequency value. Points are means across $n \geq 5$ independently sampled target regions; error bars are smaller than the plotted symbols where not visible, and lines are guides to the eye. Vertical lines mark the 431 nm Rayleigh reference, the 353 nm incoherent periodic cutoff, and the 253 nm tested period. The horizontal 10% line is a visual guide. The 253 nm point reports empirical modulation and is not a universal resolution criterion. **e**, Representative widefield and SOLIS images of 375 nm and 270 nm line-pair targets, with profiles across the 270 nm target. The 253 nm contrast measurement is shown in **d**. Scale bar, 500 nm.

## Experimental precision and temporal-sampling assessment of dynamic waveform reconstruction

Because each SOLIS map is assembled from 25 sequentially acquired subframes, we assessed experimental reconstruction precision across measured beat phases, waveform records, and operating conditions (Fig. 3a,b,d–f). A separate acquisition-level simulation evaluated the temporal sampling of a representative tip trajectory (Fig. 3c).

During the fast effective stroke, sequential acquisition produced visible differences between the cycle-averaged ciliary object and the reconstructed intensity distribution (Fig. 3b). Nevertheless, recovered tip centroids closely followed the noise-free centroid reference averaged over the same 3.58 ms acquisition windows, while a temporal-sampling control at 24 fps provided only sparse coverage of the loop (Fig. 3c). The 24 fps points were generated by temporal downsampling of the same simulated trajectory.

Across 22 beat phases and 36 repeated experimental measurements per phase, the median Euclidean deviation from the phase-specific experimental reference remained below one reconstructed pixel (87 nm), although the 2.5th–97.5th percentile interval broadened during fast phases (Fig. 3d). We therefore report phase-resolved experimental localisation precision rather than accuracy against an external ground truth. Across 80 experimental waveform records, recovered CBF, stroke amplitude, loop area, and effective-stroke fraction were compared with their corresponding experimental reference values (Fig. 3e). Median relative errors were 0.8% for CBF, 4.0% for stroke amplitude, and 9.7% for loop area; the median absolute deviation in effective-stroke fraction was 0.03. These measurements quantify the experimental precision of waveform-derived metrics across the observed range.

Experimental localisation precision was then mapped against imaging SNR and maximum within-cycle tip displacement (Fig. 3f). Median deviation increased at lower SNR and greater motion, while the typical operating condition (approximately 1.0 μm, SNR approximately 10) remained within the subpixel region. The contours provide guides across the sampled experimental conditions. Together, the experimental measurements establish subpixel tip-centroid precision over the tested phases and typical operating range. The simulation in Fig. 3c separately illustrates the temporal-sampling requirement for trajectory recovery.

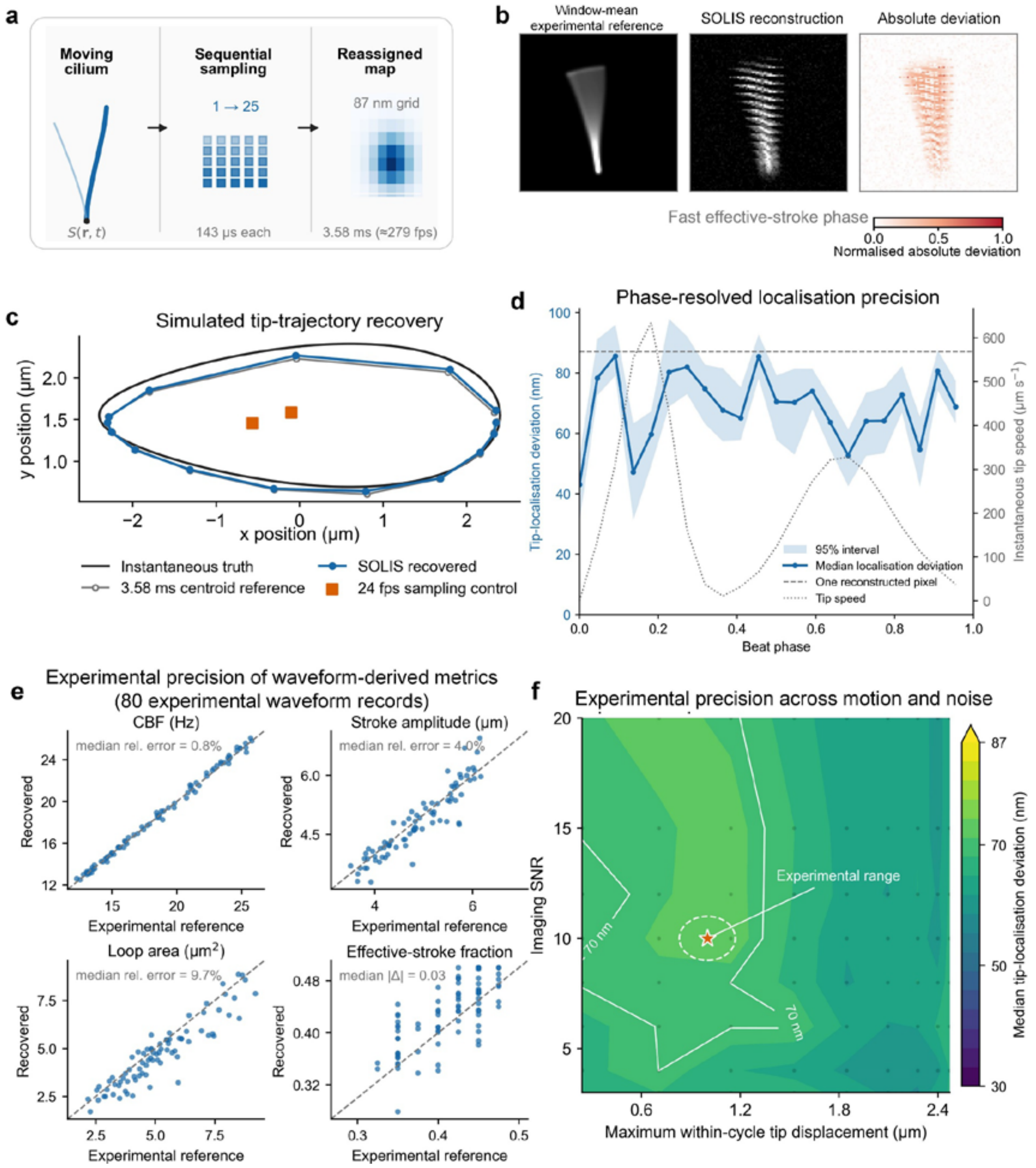

**Fig. 3. Experimental precision and temporal-sampling assessment of dynamic ciliary waveform reconstruction. a**, A time-dependent ciliary object $S(\mathbf{r},t)$ is sampled by 25 sequential DMD addresses, each integrated for 143 µs, and reassigned to an 87 nm grid to produce one map every 3.58 ms (approximately 279 reconstructed fps). **b**, Representative experimental fast effective-stroke phase. From left to right: window-mean experimental reference, SOLIS reconstruction, and normalised absolute deviation between them. The deviation is shown on a common normalised intensity scale; scale bar, 1 µm. **c**, Simulated tip-trajectory recovery. Black, instantaneous trajectory; grey open circles, centroids averaged over each 3.58 ms acquisition window; blue circles, recovered SOLIS centroids; orange squares, the same trajectory sampled

at 24 fps. **d**, Phase-resolved experimental localisation precision. The blue line and shaded band are the median and 2.5th–97.5th percentile interval of Euclidean deviations from the phase-specific experimental reference across 36 repeated measurements at each of 22 phases; the dashed line marks one 87 nm reconstructed pixel, and the dotted curve is tip speed. **e**, Experimental precision of waveform-derived metrics across 80 waveform records. Points compare recovered values with the corresponding experimental reference values; dashed lines indicate identity. Text gives median relative error for CBF, stroke amplitude, and loop area and median absolute deviation for effective-stroke fraction. **f**, Experimental localisation precision over measured SNR–motion conditions. Dots mark sampled conditions, contours interpolate median deviation, the star marks the typical operating condition, and the dashed ellipse indicates its observed range. Values >87 nm are saturated. Panel (**c**) is a simulation; panels (**b**) and (**d**)–(**f**) report experimental measurements. Experimental deviations are calculated relative to the corresponding internal experimental references and therefore quantify precision rather than accuracy against an external ground truth.

## Label-free structural discrimination and millisecond-scale imaging of motile cilia

As a secondary test of fluorescence compatibility, SOLIS distinguished mitochondrial tubules separated by 142 nm in fixed MitoTracker-labelled BPAE cells (Supplementary Fig. 4). In live COS-7 cells, the fluorescence photon budget required approximately 78 ms per address and 1.95 s per reconstructed map; SOLIS followed mitochondrial remodelling for 180 s while retaining 82% of the initial signal (Supplementary Fig. 5 and Supplementary Movie 1). This measurement characterises signal retention under the stated acquisition conditions. Phototoxicity was not assessed.

We then applied label-free SOLIS to actively beating cilia in differentiated primary human nasal epithelial cultures (NA = 0.75, λ = 530 nm). In the same region, the diagonal widefield–SOLIS comparison shows features that merge in the widefield view but remain distinguishable after SOLIS reconstruction (Fig. 4a,b).

Along the indicated profile, SOLIS separated adjacent maxima 346 nm apart, below the 431 nm Rayleigh two-point reference. The widefield profile contained one merged response (Fig. 4c). A further modulation at approximately 195 nm is visible but is not assigned as an independently resolved feature because this spacing is below the validated label-free two-point regime. A parameterised sidelobe control excludes a simple two-line sidelobe explanation under the stated probe model but does not independently establish a third physical cilium (Supplementary Note 2 and Supplementary Fig. 6).

Sequential reconstructions at 3.58 ms intervals densely sampled the beat (Fig. 4d and Supplementary Movie 2). Each map integrates one complete 25-pattern scan. The comparison frame labelled 24 fps was acquired with a 20 ms exposure; the frame interval is approximately 41.7 ms. The principal live-cilia figure of merit combines label-free sub-diffraction discrimination with millisecond-scale kinematic sampling. Each map remains an acquisition-window average during rapid motion.

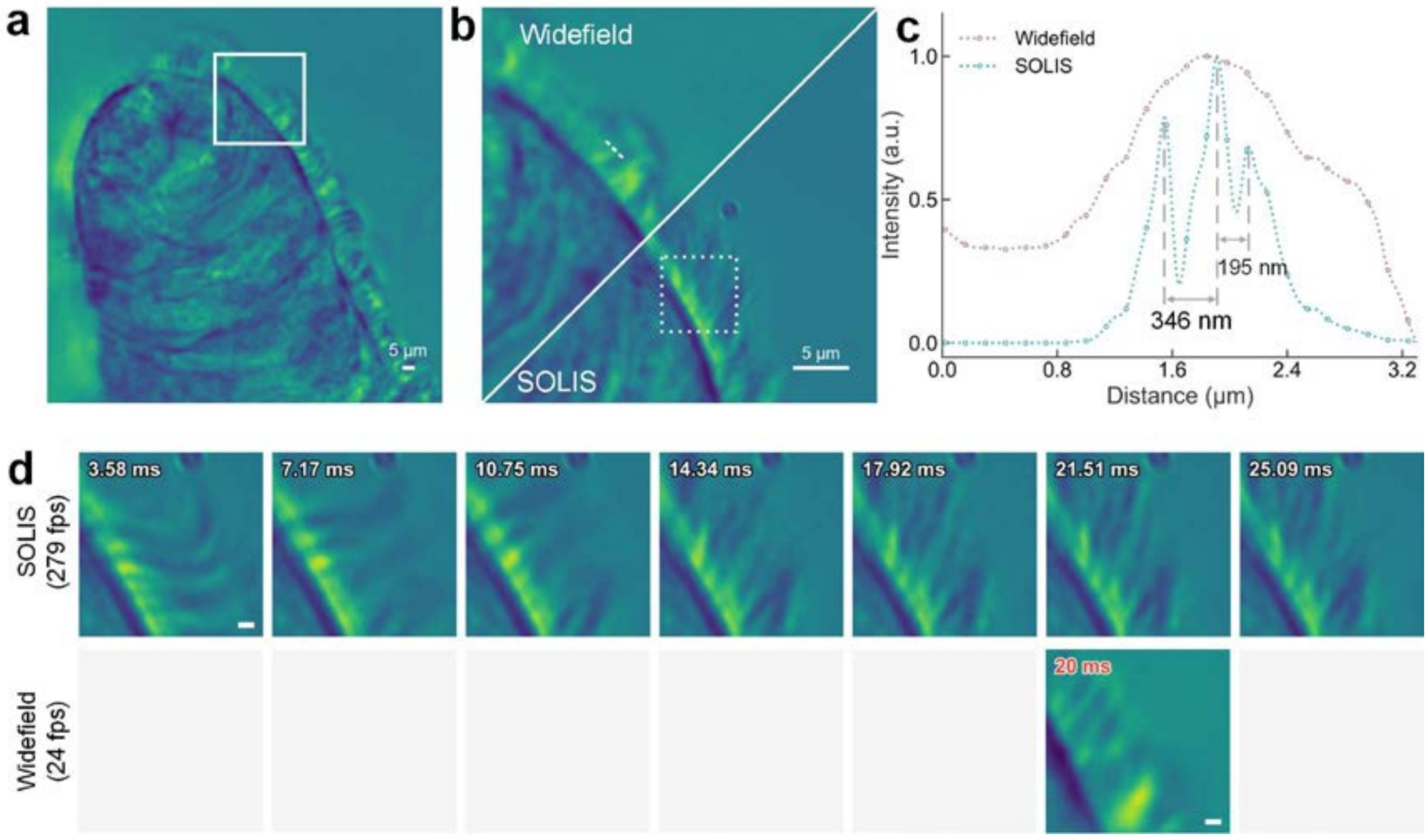

**Fig. 4. Label-free sub-diffraction discrimination and millisecond-scale imaging of cilia in differentiated primary human nasal epithelial cultures. a**, Label-free widefield image; the box marks the region enlarged in (**b**). **b**, Diagonally split comparison of the same region (widefield, upper left; SOLIS, lower right) acquired with NA = 0.75 and λ = 530 nm. The dashed line marks the profile in c, and the dotted box marks the sequence in (**d**). **c**, Normalised profiles along the line in (**b**). SOLIS distinguishes maxima separated by 346 nm, below the 431 nm Rayleigh reference. The maxima merge in the widefield profile. The additional modulation at approximately

195 nm is not assigned as an independently resolved feature; its sensitivity to sidelobes is assessed in Supplementary Note 2 and Supplementary Fig. 6. **d**, SOLIS maps at 3.58 ms intervals; each integrates one 25-pattern acquisition. The lower image is one 24 fps widefield frame acquired with a 20 ms exposure. The complete sequence is provided in Supplementary Movie 2. Scale bars, 5 µm (**a**,**b**) and 1 µm (**d**).

### Reconstruction and population-level reproducibility of ciliary waveforms

Densely sampled SOLIS centroids reconstructed a closed, asymmetric tip trajectory for the representative cilium identified in Fig. 4b (Fig. 5a). The coloured points encode temporal progression through the measured cycle, and the grey B-spline is a visual guide. A matched widefield sequence acquired independently at 24 fps yielded two tip-centroid samples over the displayed cycle and did not recover the closed loop. This comparison illustrates temporal sampling under the stated acquisition conditions and is not a downsampled SOLIS trajectory or a benchmark against dedicated high-speed widefield microscopy.

Phase folding and trajectory alignment resolved the rapid effective stroke and slower recovery stroke (Fig. 5b and Supplementary Movie 3). Each centroid is an acquisition-window–averaged position. The experimental precision analyses in Fig. 3d–f quantify phase-resolved and waveform-level variability within the measured operating range, while Fig. 3c illustrates the temporal-sampling requirement.

Across 20 tracked cilia, individual phase-averaged waveforms retained the asymmetric loop topology (Fig. 5c). The population mean is shown as red and blue segments for the effective and recovery strokes, respectively; the orange trace identifies the representative cilium in Fig. 5b. Cilia were subsamples nested within five differentiated culture dishes, not independent biological replicates.

Effective stroke fraction was below 0.5 for all 20 cilia (median, 0.34; IQR, 0.32–0.38; Fig. 5d). Per-cilium cycle-to-cycle tip localisation precision was 63.5 nm (IQR, 52.5–66.5 nm; range, 44–74 nm), and every per-cilium summary remained below one reconstructed pixel (87 nm; Fig. 5e). CBF was 19.0 Hz (IQR, 18.6–19.8 Hz), stroke amplitude was 4.94 µm (IQR, 4.55–5.38 µm), and normalised loop area was 0.70 (IQR, 0.68–0.72) (Supplementary Note 3,

Supplementary Fig. 7, and Supplementary Table 1).

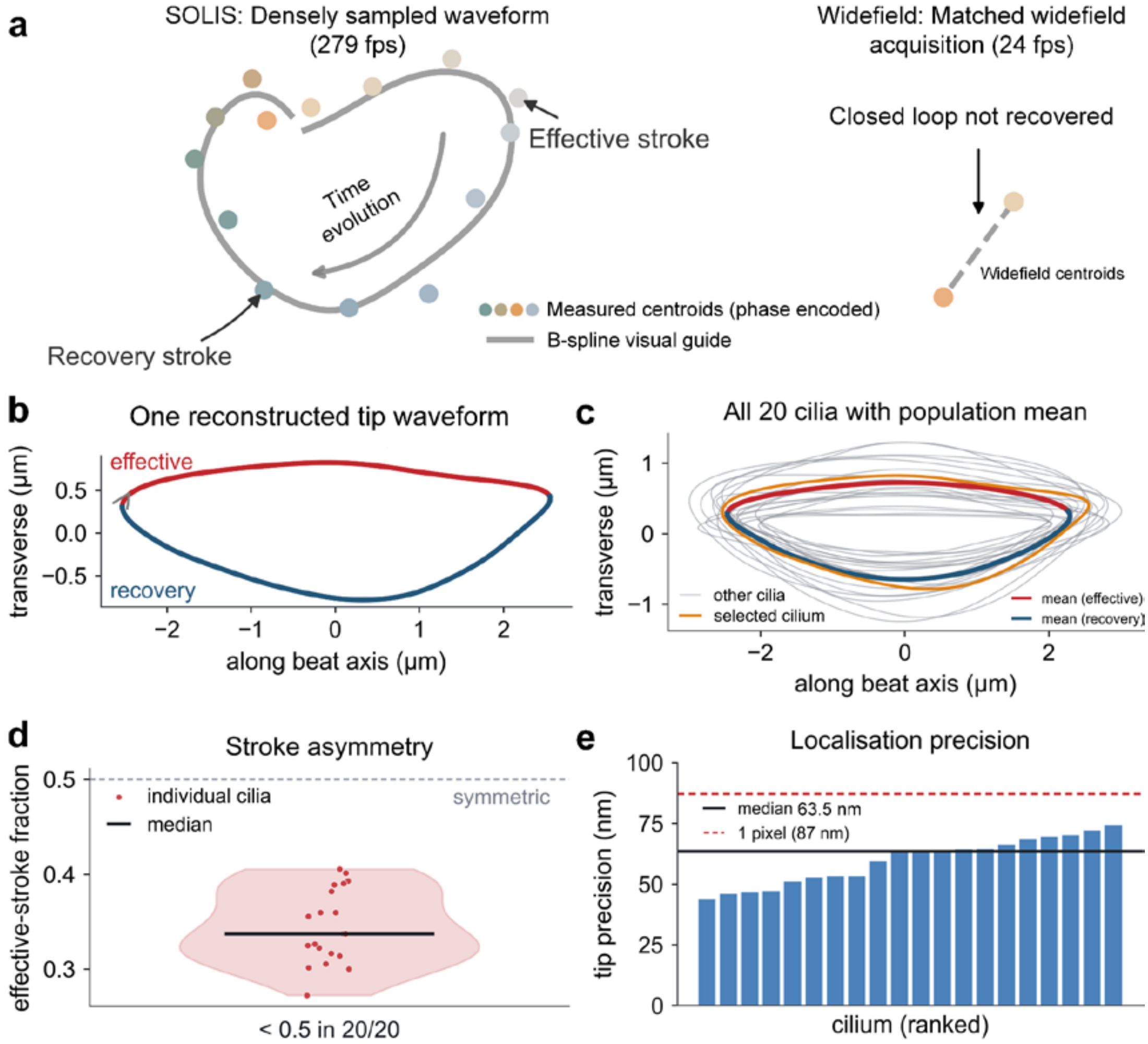

**Fig. 5. Reconstruction and population-level reproducibility of human ciliary waveforms.** **a**, Tip-centroid trajectory of the representative cilium in Fig. 4b. Coloured points are measured SOLIS centroids in temporal order and encode beat phase; the grey B-spline is a visual guide. The right-hand panel shows tip centroids extracted from a matched widefield sequence acquired independently at 24 fps (20 ms exposure; frame interval, approximately 41.7 ms). **b**, Phase-averaged waveform showing the effective (red) and recovery (blue) strokes. **c**, Phase-aligned mean waveforms of 20 cilia. Grey thin lines, individual cilia; orange, representative cilium; bold red and blue, population-mean effective and recovery strokes. **d**, Effective stroke fraction. The pink violin is a kernel-density estimate, red points are individual cilia, the black bar is the median, and the dashed line marks 0.5. **e**, Cycle-to-cycle tip-localisation precision. Each bar is one ranked cilium; the black line marks the median (63.5 nm) and the red dashed line one reconstructed pixel (87 nm). The 20 cilia are subsamples nested within five differentiated culture dishes; no inferential population test was applied. Additional

metrics and provenance are provided in Supplementary Note 3, Supplementary Fig. 7, and Supplementary Table 1.

### SOLIS-derived kinematics detect pharmacological suppression of ciliary motility

We tested whether SOLIS-derived kinematics detect a defined motility perturbation by treating differentiated cultures with the dynein inhibitor Dynarrestin (2.5 μM, 24 h; Fig. 6 and Supplementary Movie 4). The culture dish was the experimental unit. The analysis included five vehicle-control and five treated dishes. Fields of view and ciliary clusters were treated as nested subsamples.

At the culture level, mean CBF decreased from 19.2 ± 0.6 Hz in control cultures to 6.6 ± 0.5 Hz after Dynarrestin, and mean stroke amplitude decreased from 4.8 ± 0.3 to 2.3 ± 0.2 μm (mean ± s.d.; two-sided exact Mann–Whitney U test, $U = 25$, $P = 0.0079$ for each outcome). All five treated-culture means were below the control range. Cluster histograms and density maps visualise within-culture heterogeneity only; the P value displayed in Fig. 6b is explicitly labelled as a culture-level result.

Phase-averaged tip-waveform loops were smaller after Dynarrestin (Fig. 6d), and the median enclosed trajectory area decreased from 5.6 to 0.45 $\mu m^2$ across eight sampled cilia per condition (Fig. 6e). These cilium-level distributions are descriptive. The experiment is a proof of principle at one concentration and time point and does not establish dose response, reversibility, donor-level generalisability, or diagnostic performance.

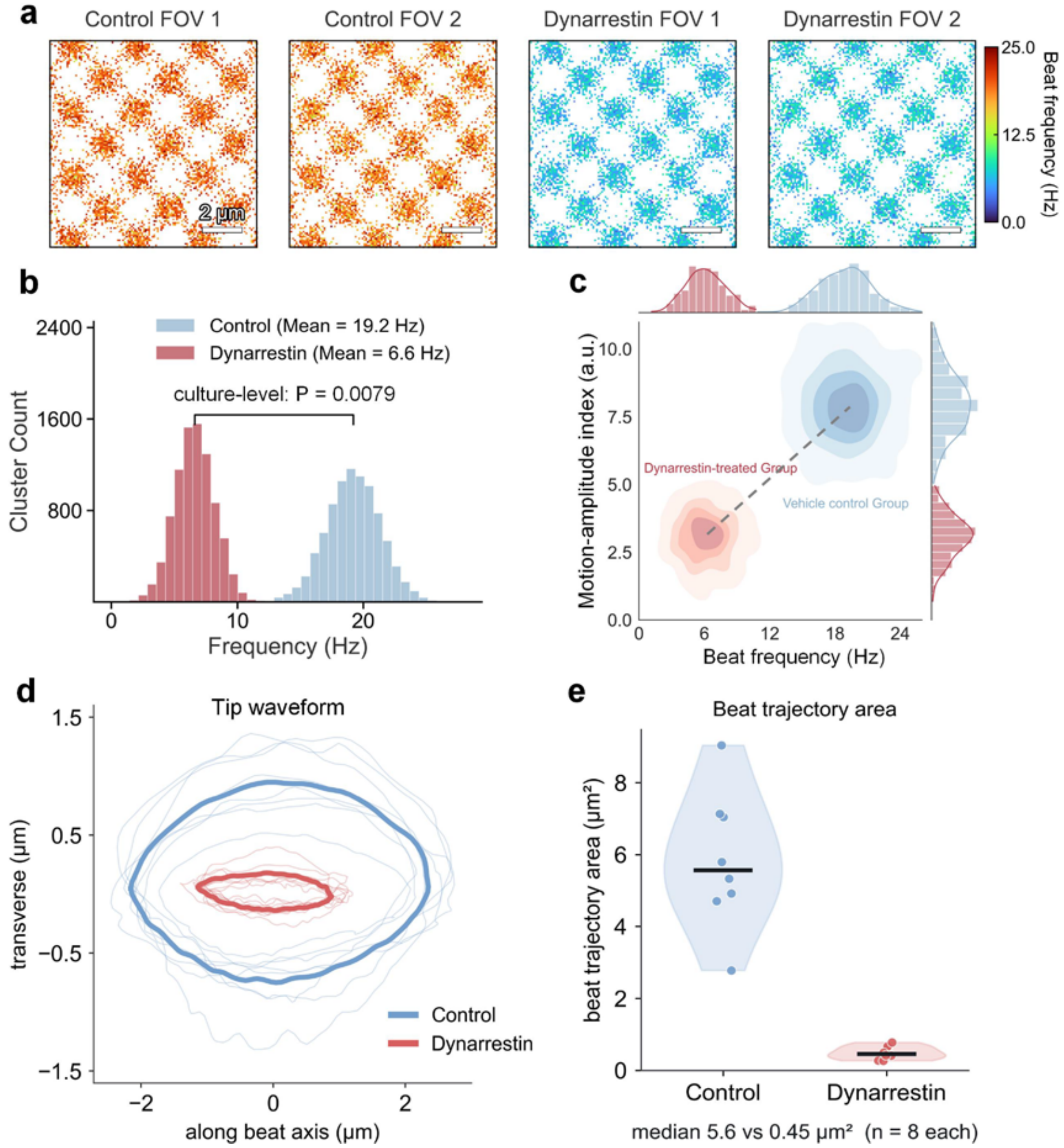

**Fig. 6. SOLIS detects Dynarrestin-induced suppression of ciliary motility. a**, Representative per-pixel ciliary beat frequency (CBF) maps from two vehicle-control and two Dynarrestin-treated fields of view, displayed on a shared colour scale. **b**, Pooled cluster-level CBF distributions. Bars show cluster counts, whereas the indicated *P* value was calculated from culture-dish means ($n = 5$ dishes per condition) and does not represent a cluster-level test. **c**, Joint cluster-level distributions of beat frequency and the motion-amplitude index defined in Methods (a.u.). Shaded contours show two-dimensional kernel-density estimates, marginal plots show the corresponding one-dimensional distributions, and the dashed line connects the group-density centres. **d**, Phase-averaged tip trajectories of eight sampled cilia per condition. Thin lines represent individual cilia, and bold lines represent group means. **e**, Enclosed trajectory areas for the same cilia. Points represent individual cilia, violins show kernel-

density estimates, and horizontal bars indicate medians. Cluster-level distributions in (**b**) and (**c**) and cilium-level measurements in (**d**) and (**e**) are descriptive nested-subsample analyses and were not treated as independent biological replicates. At the culture-dish level, CBF was 19.2 ± 0.6 Hz in control cultures and 6.6 ± 0.5 Hz after Dynarrestin treatment, whereas stroke amplitude was 4.8 ± 0.3 and 2.3 ± 0.2 μm, respectively (mean ± s.d.; two-sided exact Mann–Whitney $U$ test, $U$ = 25 and $P$ = 0.0079 for both outcomes; $n$ = 5 dishes per condition). Scale bar, 2 μm (a).

## Discussion

Representative operating regimes and their principal trade-offs are summarised in Table 1. Structured illumination microscopy and stimulated-emission-depletion microscopy provide established fluorescence-super-resolution benchmarks [33–36]. SOLIS targets label-free ciliary kinematics with millisecond-scale reconstructed sampling. Fluorescence nanoscopy provides finer resolution in labelled specimens, whereas dedicated high-speed widefield or phase-contrast microscopy provides higher raw frame rates for beat frequency analysis. SOLIS instead combines label-free contrast, sub-diffraction lateral discrimination, and millisecond-scale reconstructed sampling in differentiated primary human nasal epithelial cultures. Its relevant figure of merit is this joint operating regime, not superiority on any single axis.

Relative to our previous scan-free fluorescence implementation [20], SOLIS changes both the physical bottleneck and the biological timescale. The earlier liquid-crystal SLM required approximately 50 ms per pattern. Here, static DOE field compression is combined with binary DMD addressing at 143 μs per pattern, giving a 3.58 ms 25-position scan. This advances the architecture from labelled, slowly varying structural imaging to label-free kinematic reconstruction of fast ciliary motion. Finite DMD switching remains, but translated-stage acceleration and settling are removed.

The fluorescence experiments define a secondary operating envelope. They distinguish a 142 nm mitochondrial separation in fixed cells and follow live-cell remodelling with a 1.95 s reconstructed-map interval (Supplementary Figs. 4 and 5). The 82% signal retention over 180 s reports photostability under the stated conditions. Phototoxicity was not assessed, and the experiment was not designed as a benchmark against established fluorescence super-resolution methods.

The principal biological demonstration is the label-free ciliary measurement in Figs. 4 and 5. Millisecond-scale SOLIS sampling recovered a closed asymmetric tip trajectory, whereas the independently acquired matched widefield sequence at 24 fps yielded only two phase samples over the displayed cycle and did not recover the loop. Dynarrestin treatment shows analytical sensitivity to a strong motility perturbation at the culture-dish level. Because donor-level independence was unavailable and only one dose and time point were tested, these results do not establish biological generalisability, disease specificity, or diagnostic thresholds.

The current implementation has several limitations. Super-oscillatory illumination redistributes energy into sidelobes, and the 5×5 array is a throughput–background compromise. Label-free intensity reassignment is sensitive to specimen-dependent contrast, multiple scattering, and out-of-plane motion. The present implementation recovers a two-dimensional projection; axial excursion was not measured and can therefore not be asserted to lie within the simulated tolerance. Each map integrates 25 subframes over 3.58 ms and must be interpreted as an acquisition-window average. The sampling analysis in Fig. 3c uses an analytic probe, nominal gain distribution, and the experimentally implemented address order and is reported separately from the experimental precision measurements in Fig. 3b,d–f.

Future implementations could incorporate measured, address-specific probe kernels and gains, the actual DMD acquisition order, multiplane detection, and automated trajectory analysis across donor-characterised cohorts. The central architectural result is that spatial compression and temporal addressing need not be performed by the same physical mechanism. SOLIS complements fluorescence super-resolution and diffraction-limited high-speed ciliary imaging by opening a label-free regime that combines sub-diffraction discrimination with millisecond-scale kinematic sampling.

## Methods

### Optical implementation and SOLIS system configuration

The SOLIS system used separate illumination configurations for label-free and fluorescence imaging. Label-free targets and cilia were illuminated with a 530 nm LED (M530L4, Thorlabs) and a 10 nm bandpass filter (FBH530-10, Thorlabs). The collimated, spatially filtered beam provided adequate spatial coherence across the DOE aperture, as verified by the measured focal profiles in Fig. 2c, while the finite bandwidth reduced laser-like speckle. Fluorescence imaging used 561 nm continuous-wave excitation. In both configurations, a plano-convex lens (L1, f = 200 mm) expanded the beam to fill the active DMD aperture.

The expanded beam was directed onto a high-speed DMD (V-6501 VIS, ViALUX) through a total-internal-reflection prism (VSHV, Young Optics) at an incident angle of 73°, giving an effective pixel pitch of 7.95 μm in the scan direction relative to the 7.60 μm physical mirror pitch. The DMD encoded programmable binary diffraction holograms. The selected first diffraction order was isolated with an iris diaphragm (ID50, Thorlabs) at the back focal plane of relay lens L3 (f = 300 mm), consistent with Fig. 2a and Supplementary Fig. 2a. A multilevel phase-only DOE was placed in the Fourier plane of the illumination relay, conjugate to the objective back aperture, to generate the super-oscillatory focal array at the sample plane.

The detection path was configured for the photon budget and spatial-performance requirements of each modality. Ciliary imaging used a 0.75 NA dry objective (20×, Nikon) with 530 nm illumination. Fluorescence imaging used a 1.45 NA oil-immersion objective (Nikon CFI Apochromat) with 561 nm excitation; emitted light was isolated with a Nikon B-2A filter set. Images were recorded with an ORCA-Fire CMOS camera (C16240-20UP, Hamamatsu) through CoaXPress. High-speed acquisition used a 30×30-pixel subarray and a 143 μs exposure per subframe, matched to dwell time $\tau$, corresponding to approximately 7 kHz raw subframe acquisition and a 25-subframe, 3.58 ms scan cycle. The manufacturer-specified 19,500 fps four-line and 115 fps full-frame rates are camera readout limits and are distinct from the rate of complete SOLIS maps.

DMD patterns and camera exposures were synchronised by TTL triggers from a Python control suite. One label-free ciliary scan comprised 25 sequential binary holograms, each integrated for 143 μs, for a measured total of 3.58 ms and 279 reconstructed fps. This is the rate of complete reassigned maps, not the raw mirror-switching rate. In fluorescence mode, each address was integrated for approximately 78 ms, giving 1.95 s per 25-position reconstructed map. Pattern generation, triggering, acquisition, and hardware coordination were automated by the same control suite.

**DOE design, fabrication and optical probe characterisation**

The super-oscillatory probes were generated by phase-only DOEs, one optimised for each configuration, a 530 nm / 0.75 NA element (label-free) and a 561 nm / 1.45 NA element (fluorescence). The DOE was designed to shape the angular spectrum of the incident beam and generate either a single compressed focal spot or a 5×5 array of sub-diffraction focal spots at the sample plane. The high-NA element was optimised with vectorial diffraction theory; the 0.75 NA element used the scalar approximation.

DOE phase profiles were optimised with a modified iterative Fourier transform algorithm (IFTA; Supplementary Fig. 9). Gradual zero-padding refined focal-plane sampling from $\lambda f/D$ to $\lambda f/(10D)$, and weighted amplitude constraints balanced central-lobe width, optical efficiency, and sidelobe background. Sixteen-level phase quantisation was applied during optimisation; Supplementary Fig. 9b shows convergence of the design-stage signal-window efficiency.

The 16-level phase-only DOEs were fabricated on 30×30×1 mm fused-silica substrates. A 2.3 µm positive photoresist layer (AZ6130, Merck Millipore) was spin-coated at 2000 rpm and patterned by contact UV lithography using a mask aligner (SÜSS MicroTec MA6/BA6; i-line, 365 nm; nominal resolution ≈1 µm; alignment accuracy ≈0.5 µm). The exposed pattern was transferred into silica by reactive-ion etching (STS M/PLEX ICP-AOE; $O_2/C_4F_8$, ≈10 Å $s^{-1}$). Four aligned binary lithography-and-etch steps were used to generate the 16-level phase profile (Fig. 2a, inset). Atomic-force microscopy gave groove depths of 583 nm for the label-free DOE designed for 530 nm illumination and 621 nm for the fluorescence DOE designed for 561 nm illumination, close to the theoretical values of 588 and 623 nm, respectively. These correspond to phase errors below 1%, with a theoretical 16-level diffraction efficiency of 98.7%. The resolution target was a nanolithographic line-pair pattern with sub-diffraction features.

Focal-plane intensity distributions were background-subtracted and analysed from normalised central-lobe cross-sections. For the label-free element, the central-lobe FWHM was 203 nm for one focus and 224 nm for the 5×5 array, compared with a 360 nm diffraction-limited FWHM. Empirical contrast-transfer analysis retained measurable contrast at 253 nm (Fig. 2d), whereas representative image-based resolution was demonstrated at 270 nm (Fig. 2e). The 561 nm/1.45 NA fluorescence element distinguished a 142 nm mitochondrial separation (Supplementary Fig. 4c), below the 236 nm Rayleigh reference. The 5×5 optical efficiency was approximately 30%, defined as power delivered to the target focal array after spatial filtering divided by incident power.

For each array configuration, photometric uniformity was quantified from the peak intensities of the active focal spots. Intensity non-uniformity was defined as the coefficient of variation of peak intensity [37],

$$\text{Non-uniformity} = \frac{\sigma(I_{\text{peak}})}{\mu(I_{\text{peak}})} \times 100\% \ , \tag{3}$$

where $I_{\text{peak}}$ is the peak intensity of focal spot, and $\sigma$ and $\mu$ are the standard deviation and mean across all active spots. SBR was defined as the central-lobe peak intensity divided by the mean intensity within an annulus spanning 1.0–2.0 Rayleigh units from the spot centroid, where 1 Rayleigh unit = 0.61λ/NA. This region captures first-order sidelobe and interspot background contributions. Probe-characterisation measurements used n ≥ 5 independent acquisitions under matched illumination, exposure, and camera-gain conditions; values are reported as mean ± s.e.m. unless stated otherwise.

For contrast-transfer analysis (Fig. 2d), $C = (I_{\max} - I_{\min})/(I_{\max} + I_{\min})$ was measured from nanofabricated targets with periods of approximately 1000–200 nm under matched widefield and SOLIS acquisition. Each point was averaged over n ≥ 5 independently sampled target regions and normalised to the lowest spatial frequency contrast; standard errors were smaller than the symbols where not visible.

### Digital hologram generation and scanning calibration

Lateral displacement of the super-oscillatory probe array was achieved by modulating the carrier spatial frequency of the binary diffraction hologram encoded on the DMD. Each of the 25 binary holograms encoded a linear phase ramp (a binary grating) whose carrier spatial frequency placed the first diffraction order at a defined position in the relay Fourier plane, displacing the super-oscillatory probe array to the corresponding sample-plane position $\mathbf{d}_n$ (Eq. (2)).

The scan interval was selected to oversample the 224 nm central lobe. Centroid shifts across consecutive holograms gave $\Delta x = \Delta y = 87$ nm in the sample plane (Fig. 2b). The 25 calibrated coordinates formed one regular 5×5 unit cell. The maximum digital steering range was 333 μm, set by the smallest DMD carrier period that could be encoded reliably at the operating wavelength and relay magnification.

The optical arrangement and a conceptual position–time comparison with an inertial scanner are shown in Supplementary Fig. 2. The schematic illustrates the absence of mechanically induced acceleration and settling in digital addressing; it is not a time-resolved switching measurement and does not imply zero micromirror transition time.

### Experimental precision analysis and temporal-sampling simulation

Experimental precision analyses for Fig. 3b,d–f used the same 25-address reconstruction and 87 nm grid as the ciliary measurements. Figure 3b compares a representative reconstruction with its window-mean experimental reference from the same 3.58 ms interval. Phase-resolved precision in Fig. 3d was calculated at 22 phases from 36 repeated experimental measurements per phase as the Euclidean deviation from the phase-specific experimental reference centroid.

Figure 3e analysed 80 experimental waveform records with the same centroid extraction and waveform-metric pipeline used elsewhere. CBF, stroke amplitude, normalised loop area, and effective-stroke fraction were compared with the corresponding reference values derived from each cycle-averaged experimental waveform. Relative error was reported for CBF, stroke amplitude, and loop area; absolute deviation was used for effective-stroke fraction.

Figure 3f grouped experimental localisation deviations by imaging SNR and maximum within-cycle tip displacement. The median phase-resolved Euclidean deviation was calculated at each sampled condition, and contours interpolate between measured conditions. Values above 87 nm were saturated on the colour scale. The star marks the typical experimental condition, and the dashed ellipse indicates its observed range.

Only Fig. 3c used simulation. A planar cilium was represented by a deforming cubic Bézier centreline with a nominal CBF of 19.2 Hz, a principal-axis stroke amplitude of 4.8 μm, and an effective-stroke fraction of 0.34. Each cycle comprised 25 sequential 143 μs exposures, giving a 3.575 ms acquisition window reported as 3.58 ms, and measurements were reassigned to a 96×96 grid sampled at 87 nm. The analytic probe comprised a 224 nm FWHM Gaussian central lobe and a ring at 0.47 μm with a peak ratio of 1/5.3. Nominal gain variation was 7.1%, residual flat-field variation was 1%, background was 12 electrons, read-noise s.d. was 2.2 electrons, nominal SNR was 10, and the address sequence was serpentine. This simulation illustrates temporal sampling and is separate from the experimental precision analyses in Fig. 3b,d–f.

### Human subjects, cell culture and sample preparation

**Human-derived nasal epithelial cell cultures and ethics.** Primary cells were obtained from de-identified nasal-brushing specimens collected at Beijing Children's Hospital, Capital Medical University. The imaging investigators received de-identified preparations without donor identifiers or linked clinical metadata and performed subsequent expansion, air–liquid-interface differentiation, and imaging. Use of the cells was approved by the Ethics Committee of Beijing Children's Hospital, Capital Medical University (approval no. [2024]-E-132-R), in accordance with the Declaration of Helsinki. Written informed consent was obtained from donors or, for minors, from their parents or legal guardians. Ten differentiated culture dishes were used in the perturbation analysis (five vehicle control and five Dynarrestin treated). Because donor-level independence could not be verified, the dish was the experimental unit and fields and clusters were nested subsamples.

**BPAE cell preparation and COS-7 cell culture.** Commercially prepared fixed bovine pulmonary artery endothelial (BPAE) cells pre-labelled with MitoTracker Red CMXRos were obtained as a fluorescence microscopy reference slide (FluoCells Prepared Slide #2, F14781, Thermo Fisher Scientific) and used directly for fluorescence resolution validation without further processing. COS-7 cells were cultured in Dulbecco's modified Eagle medium (DMEM, Gibco) supplemented with 10% foetal bovine serum (FBS, Gibco) and 1% penicillin–streptomycin (Gibco). Cells were maintained in a humidified incubator at 37 °C with 5% $CO_2$. For live-cell SOLIS imaging, COS-7 cells were seeded onto 35 mm glass-bottom dishes (No. 1.5H coverslip, MatTek) at densities selected to achieve 60-70% confluency at the time of acquisition. To label mitochondria, live COS-7 cells were incubated with MitoTracker Red CMXRos (100 nM, Invitrogen) for 20 min at 37 °C, washed twice with pre-warmed phosphate-buffered saline (PBS), and transferred to phenol-red-free Live Cell Imaging Solution (Invitrogen) before acquisition.

**Preparation of primary human nasal epithelial cells.** Primary human nasal epithelial cells obtained from nasal brushing were expanded as basal cells in PneumaCult-Ex Plus medium (STEMCELL Technologies) on collagen I-coated standard tissue culture flasks under submerged conditions. Upon reaching 80–90% confluency, basal cells were dissociated with Animal Component-Free Cell Dissociation Kit (STEMCELL Technologies), counted, and seeded onto collagen I-coated Transwell permeable supports (0.4 µm pore size, 24-well plate, Corning) at a density of approximately $3.3\times10^4$ cells per insert. After cells reached 100% confluency on the Transwell membrane (approximately 3-5 days), the apical medium was removed to establish an air–liquid interface, and the basal medium was switched to PneumaCult-ALI medium (STEMCELL Technologies). Cultures were maintained at ALI for 28 days with basal medium changes every 2–3 days to promote full ciliogenesis and differentiation [38, 39]. Active ciliary beating was confirmed by brightfield microscopy before SOLIS acquisition. Ciliary clusters were selected for analysis based on multiciliated morphology and sustained oscillatory activity. Non-motile regions, sparsely ciliated areas and regions with obvious debris, focal drift or loss of focus were excluded before quantitative analysis.

For pharmacological perturbation experiments, Dynarrestin (MedChemExpress; catalogue no. 121802) was prepared as a DMSO stock and added to the culture medium at 2.5 µM (1 µl stock per 1 ml medium; 0.1% v/v DMSO) for 24 h at 37 °C. Vehicle-control cultures received the same volume of DMSO. Cultures were imaged under matched optical and environmental conditions after treatment.

**Imaging environment.** Live-cell and ciliary imaging was performed at 37 °C in a stage-top chamber. COS-7 cells were maintained in phenol-red-free Live Cell Imaging Solution (Invitrogen). Differentiated nasal epithelial cultures were maintained in pre-warmed Dulbecco's phosphate-buffered saline (DPBS, Gibco; pH 7.4) during short acquisitions.

**Image reconstruction, preprocessing and intensity-structure mapping**

Each SOLIS cycle produced 25 raw intensity frames, one per addressed hologram. Frames were mapped to calibrated sample-plane coordinates after background subtraction, flat-field correction, subframe registration, lattice assembly, and intensity normalisation.

Reconstruction was calibrated spatial reassignment of measured scattering-contrast intensities; it did not use deconvolution, regularisation, learning-based restoration, or synthesis of unmeasured structure. The full processing sequence and matched profiles are provided in Supplementary Fig. 3.

Preprocessed intensities were assigned to the corresponding 87 nm reconstructed coordinates. Spatial performance was determined empirically. The central-lobe FWHM was 224 nm, contrast remained measurable at a 253 nm target period, and representative target modulation was resolved at 270 nm (Fig. 2d,e). The mapping assumes a locally planar, weak-contrast specimen over one scan. Actual axial ciliary excursion was not measured. The axial simulation in Supplementary Note 1 defines a conditional tolerance and does not validate individual trajectories. Dynamic centroid error is assessed separately in Fig. 3.

No averaging was applied across reconstructed maps; each 25-position cycle was processed independently. A fixed Gaussian kernel with $\sigma \approx 50$ nm (approximately 0.6 reconstructed pixel) attenuated residual high-spatial-frequency background. Because this width is below one third of the 224 nm central-lobe FWHM, it was not used to set the nominal spatial performance scale. The final label-free reconstruction and its comparison with widefield imaging are shown in Supplementary Fig. 3e–g. The effect of Gaussian filtering was not evaluated as a separate processing step.

For ciliary analysis, distal tip centroids were estimated in a local region by two-dimensional Gaussian fitting. Twenty cilia with identifiable tips that remained in focus for at least one complete beat were followed for the descriptive population analysis. Measured centroids were the quantitative data. A B-spline was used only in Fig. 5a as a visual guide; phase-averaged experimental waveforms were obtained by Fourier fitting as described in Supplementary Note 3.

**Ciliary waveform reconstruction and kinematic analysis**

Active ciliary clusters were segmented from the pixel-wise temporal standard deviation using a mean + 3 s.d. background threshold and connected-component filtering. High-speed acquisition used a 30×30-pixel subarray corresponding to approximately 13×13 μm and was reconstructed on the 87 nm grid. The 169×169 μm maximum field refers to lower-rate full-field operation and was not acquired concurrently with 279 reconstructed fps. Experimental SNR was defined as peak ciliary signal above local background divided by the standard deviation of that background and was approximately 10 per subframe.

CBF was estimated by applying a Hann-windowed fast Fourier transform to each cluster's intensity trace over at least five beat cycles. The dominant spectral peak within 0–50 Hz defined the pixel-level CBF. Pixels with spectral SNR < 3, calculated relative to the adjacent-band baseline, were excluded; the cluster-level CBF was the median of the remaining pixel estimates. Sub-bin peak frequency was obtained by quadratic interpolation around the spectral maximum.

Stroke amplitude was the peak-to-peak displacement along the first principal component of the phase-averaged trajectory. The principal-axis coordinate was divided at its two extrema into two contiguous trajectory segments. The segment with the greater mean absolute projected speed was designated the effective stroke and the other the recovery stroke. Effective stroke fraction was the duration of the effective segment divided by the beat period (Fig. 5b,d).

For the cluster-level visualisation in Fig. 6c, each ciliary cluster was represented by its cluster-level CBF and a motion-amplitude index derived from the temporal intensity signal. After background subtraction, the Fourier amplitude at the dominant beat frequency was calculated for each pixel assigned to the cluster and normalized to the pixel's mean temporal intensity. The cluster-level motion-amplitude index was defined as the median of these pixel-level values. The resulting dimensionless index is reported in arbitrary units and was used only as a descriptive measure of cluster-level motion. It is distinct from the physical stroke amplitude in micrometres obtained from reconstructed tip-centroid trajectories. The joint distribution for each group was estimated using two-dimensional Gaussian kernel density estimation with Scott's-rule bandwidth applied separately to each axis.

**Statistics and reproducibility**

For the pharmacological comparison, the culture dish was the unit of analysis (five control and five treated dishes; 20 fields per condition in total). Per-dish means were compared by a two-sided exact Mann–Whitney U test. Fields and clusters were nested subsamples. CBF was 19.2 ± 0.6 versus 6.6 ± 0.5 Hz and stroke amplitude was 4.8 ± 0.3 versus 2.3 ± 0.2 μm (mean ± s.d.). For each readout, all five control values exceeded all five treated values ($U = 25$, $P = 0.0079$).

Given the small number of dishes, uncertain donor-level independence, and one inhibitor dose and time point, the experiment demonstrates culture-level analytical sensitivity only (Supplementary Fig. 8). Cilia, fields, and clusters were not treated as independent biological replicates. Fluorescence compatibility experiments were repeated in at least five independent acquisitions under matched conditions. Experiments were not randomised or blinded, and no exclusions were made post hoc on the basis of effect size or outcome.

Data processing and visualisation used Python 3.13 with NumPy, SciPy, Seaborn, and Matplotlib. CBF estimation used Hann-windowed spectral analysis. Two-dimensional descriptive density maps used scipy.stats.gaussian_kde with Scott's-rule bandwidth. Custom scripts are available as stated below.

**Data availability.** Data supporting the findings are available within the article and its Supplementary Information. Source data and representative raw sequences are deposited at Zenodo (https://doi.org/10.5281/zenodo.20464998).

**Code availability.** MATLAB code for DOE design, phase estimation, and DMD hologram generation, together with Python code for reconstruction, simulation, CBF extraction, kinematic analysis, statistics, and visualisation, is available at https://github.com/ningxu1995/SOLIS and archived at Zenodo (https://doi.org/10.5281/zenodo.20464998).

## Acknowledgments

We thank Assoc. Prof. Zhendong Zhu (Peking University) for providing the custom nanofabricated resolution target and Prof. Lingfeng Pan (Institute of Semiconductors, Chinese Academy of Sciences) for fabricating the diffractive optical elements. We dedicate this work to the memory of our colleague and co-author Prof. Qiaofeng Tan, whose guidance was central to the conception and early development of this study.

## Funding

This work was supported by Cancer Research UK (C9545/A29580) and the Engineering and Physical Sciences Research Council (EP/R003599/1). This work was also supported by the National Natural Science Foundation of China (62075112).

## Author contributions

S.E.B., C.W., N.X., and Q.T. conceived the original idea. N.X. developed and designed the study, built the SOLIS imaging system, performed the experiments and data analysis, and wrote the manuscript. C.W., G.S. and Q.W. assisted with the experimental design, and Q.W. additionally performed the numerical simulations. B.X. and M.J. provided the cilia samples. G.S., M.J., and B.X. contributed to the implementation of the project. S.E.B. and C.W. supervised the project. All authors except Q.T., who passed away during this work, reviewed and approved the final manuscript.

## Competing Interests

The authors declare no competing interests.

**Table 1.** Representative operating regimes of optical modalities relevant to native ciliary waveform imaging. Values are approximate literature-reported ranges rather than head-to-head measurements. The demonstrated lateral spatial performance column contains modality-specific metrics and should not be interpreted as a uniform resolution criterion. Literature benchmarks are drawn from [10, 20, 33–36, 40–43]. SOLIS values are from the static experimental measurements in Fig. 2 and Supplementary Fig. 4 and from the high-speed acquisitions in Figs. 4–6; Fig. 3b,d–f report experimental precision, and Fig. 3c is a temporal-sampling simulation. Abbreviations: SIM, structured illumination microscopy; STED, stimulated emission depletion microscopy; SOFI, super-resolution optical fluctuation imaging; SR, super-resolution; MCoSM, mechanical-scan-free multicolour super-resolution microscopy; FOV, field of view.

| Modality | Demonstrated lateral spatial performance | Typical temporal sampling | Label-free compatibility | Scanning mechanism | Typical FOV | Main limitation for native ciliary waveform reconstruction |
| --- | --- | --- | --- | --- | --- | --- |
| High-speed widefield/phase contrast [10, 40] | ~250–500 nm | 100–1,000 fps | Yes | Single-shot, no scanning | Large | Sufficient temporal sampling but diffraction-limited structural discrimination |
| Confocal microscopy [41, 42] | ~200–300 nm | ~1–10 fps | Typically fluorescence | Galvanometer point scanning | Medium | Point scanning limits frame rate for fast waveform reconstruction |
| SIM [33, 34] | ~100–130 nm | ~1–10 fps | Typically fluorescence | Typically 9–15 raw frames per SR frame | Medium | Requires fluorescent labelling and multi-frame reconstruction |
| STED [35, 36] | ~30–80 nm | ~1 fps; sample dependent | Fluorescence | Point scanning with depletion beam overlay | Small-medium | High optical dose and scanning burden complicate sustained fast imaging |
| SOFI/fluctuation-based SR [43] | ~100–200 nm | Not real-time; effective rate ~0.01–0.1 fps | Fluorescence | Widefield acquisition requiring hundreds to thousands of frames | Medium | Temporal fluctuation statistics are not suited to direct millisecond waveform readout |
| MCoSM [20] | ~170 nm | ~20 Hz | Typically fluorescence | Electronic phase-pattern refresh, ~50 ms per pattern | Medium | Phase-modulator refresh rate limits fast periodic motion imaging |
| SOLIS (this work) | 142 nm (fluorescence); 270 nm label-free image; contrast at 253 nm | 279 reconstructed fps (label-free) | Yes | DMD addressing; no translated-stage motion | ~13×13 $\mu m^2$ at 279 reconstructed fps; ~169×169 $\mu m^2$ at lower rates | 2D window-averaged maps; axial and sequential-scan uncertainty |